\documentclass[twoside]{dis07}
\usepackage[latin1]{inputenc}
\usepackage[dvips]{graphicx,epsfig,color}
\usepackage{wrapfig,rotating}
\usepackage{amssymb,amsmath,array}

\begin{document}
\title{Status of the Forward Physics Projects in ATLAS}

\author{Stefan Ask$^1$ \\
(on behalf of the ATLAS Luminosity and Forward Physics Working Group)
%
%
\vspace{.3cm}\\
%
1- CERN - Physics Department \\
CH-1211 Geneva 23 - Switzerland
%
}

\maketitle

\begin{abstract}
The ATLAS experiment at the LHC is building several detector systems for forward physics studies and to determine the luminosity. 
The main forward systems consist of a Cerenkov detector called LUCID, a Zero Degree Calorimeter (ZDC) and Roman Pots which will 
house a scintillating fiber tracker system called ALFA. Here we report some of the forward physics activities that are foreseen in 
ATLAS together with the status of the related detector systems. 
\end{abstract}

\section{Forward detectors in ATLAS}

In addition to the main ATLAS detector, also three smaller systems are built to cover the forward region \cite{tdr_loi}. These are 
closely connected to the luminosity determination in ATLAS, but are in addition foreseen to study forward physics. When ordered by 
their distance from the ATLAS interaction point (IP) the first system is a Cerenkov detector called LUCID. LUCID is the main luminosity 
monitor in ATLAS and is located $17~m$ away from the IP. The second system is the so-called {\it zero degree calorimeter} (ZDC) which 
is located at a distance of $140~m$ from the IP. This corresponds to the location where the LHC beam-pipe is divided into two and the 
ZDC is located between the beam pipes just after the split inside the so-called TAN absorber. The most remote system is the so-called 
ALFA system. ALFA consists of scintillating fiber trackers located inside roman pots at a distance of $240~m$ from the ATLAS IP. 
All results presented below are preliminary.

ATLAS also foresee upgrades of the roman pot program with stations at $220~m$ and $420~m$ dedicated entirely to diffractive physics, 
however, the status of these projects are presented by C.~Royon \cite{rp_220} and A.~Pilkington \cite{rp_420} at this conference.

\section{The ALFA system}

The ALFA ({\it Absolute Luminosity For ATLAS}) system consists of scintillating fiber trackers located in roman pots at a distance of 
$240~m$ on each side of the IP. The roman pots allow the detectors to approach the beam inside the LHC beam-pipe and the main purpose 
of ALFA is to measure elastic proton scattering at low angles. This is primarily to determine the absolute luminosity in ATLAS, but 
also other physics studies are foreseen such as measuring the total $pp$ cross section, measuring elastic scattering parameters and 
potentially also to tag protons for diffractive studies.

For a maximum precision in the luminosity measurement, the goal is to measure elastic scattering in the Coulomb interference region, 
which requires a measurement of scattering angles down to about $3 ~\mu rad$. In order to reach such small angles, the LHC has to run 
with special so-called high $\beta^*$ optics, but even with this optics the detectors have to be located at a distance of only $1-2~mm$ 
from the beam. The main requirements on the tracker are, a spatial resolution of about $30 ~\mu m$, no significant non-active edge 
region, insensitivity to the RF from the LHC beam and to the vacuum in the roman pot. The high $\beta^*$ runs have a very low 
luminosity and for this reason no radiation hard technology have to be adopted.

Due to these requirements, ATLAS has chosen a scintillating fiber tracker. Prototype detectors of the ALFA tracker have been validated 
in beam tests at DESY \cite{alfa_nima} and CERN together with the front-end electronics and the so-called overlap detector alignment 
system. The tests have shown an adequate performance for the luminosity measurement and the full ALFA system is foreseen to be installed 
in the shutdown between 2008 and 2009.

\begin{wrapfigure}{r}{0.49\columnwidth}
\centerline{\includegraphics[width=0.47\columnwidth]{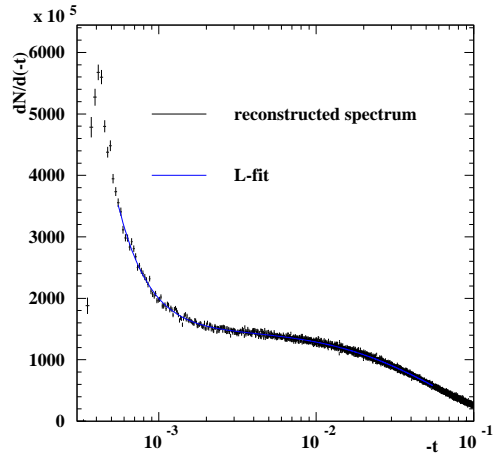}}
\caption{Simulation of the reconstructed $|t|$-distribution from the ALFA measurement.}\label{Fig:ElFit}
\end{wrapfigure}
In parallel to the detector development, the  measurement of elastic scattering have been carefully simulated. The primary analysis is 
based on a fit of the differential cross section of elastic scattering (simplified below),
\begin{equation}
\frac{dN}{dt} = L \cdot \pi \left | -\frac{2 \alpha}{|t|} + \frac{\sigma_{tot}}{4 \pi}(i + \rho)e^{-B|t|/2} \right | ^2 \nonumber
\end{equation}
\begin{equation}
t = -(p \cdot \sin{\theta})^2 \nonumber
\end{equation}
to the $t$-distribution of the data. Figure \ref{Fig:ElFit} shows the reconstructed $t$-distribution from simulations of the ALFA 
measurement. As seen in the plot, the acceptance covers the interference region where the EM contribution becomes significant and give rise 
to the steeper slope at low $t$-values. Several systematic errors have been studied, for example due to beam properties, detector acceptance, 
alignment and background. The precision of the luminosity measurement from the fit is estimated to be ${\cal L} \pm 2\% (stat) \pm 2\% (syst)$. 
Also alternative methods to determine the luminosity are foreseen such as using the optical theorem.

The fit allows a measurement of the total $pp$ cross section ($\sigma_{tot}$), the nuclear slope parameter ($B$) and the ratio of the real 
and imaginary part of the nuclear amplitude ($\rho$). The current results only includes statistical errors, but these indicates that the 
listed parameters will be possible to measure with a precision of the order of $1\%$, $0.5\%$ and $4\%$ respectivelly.

\section{The LUCID system}

LUCID ({\it LUminosity measurement using Cerenkov Integrating Detector}) is the main luminosity monitor in ATLAS. Its main purpose is to 
detect inelastic $pp$ scattering in the forward direction, both in order to measure the integrated luminosity of the ATLAS runs and for 
on-line monitoring of the instantaneous luminosity and beam conditions. Potentially LUCID could also be used for diffractive studies, e.g. 
as a rapidity gap veto.

The luminosity monitoring is based on the fact that the inelastic $pp$ rate ($R_{pp}$) seen by LUCID is proportional to the luminosity,
\begin{equation}
R_{pp} = \mu_{LUCID} \cdot f_{BC} = \sigma _{inel} \cdot \varepsilon_{LUCID} \cdot {\cal L} \nonumber
\end{equation}
Here the mean number of inelastic $pp$ interactions per bunch crossing (BC) seen by LUCID, $\mu _{LUCID}$, is related to the luminosity ($L$) 
by the inelastic cross section ($\sigma _{inel}$) and the LUCID detection efficiency ($\varepsilon_{LUCID}$). In this equation $f_{BC}$ 
represent the bunch crossing rate. The value of $\mu _{LUCID}$ can be measured by LUCID in several ways \cite{lucid_methods},
\begin{tabbing}
\= $\bullet$ Zero Counting: \hspace*{1.0cm}\=$\mu _{LUCID} = -ln(N_{ZeroBC}/N_{TotBC})$ \\
\> $\bullet$ Hit Counting:                 \>$\mu _{LUCID} = \langle N_{Hits / BC} \rangle / \langle N_{Hits / pp} \rangle $ \\
\> $\bullet$ Particle Counting:            \>$\mu _{LUCID} = \langle N_{Particles / BC} \rangle / \langle N_{Particles / pp} \rangle $. 
\end{tabbing}
The first method determines $\mu _{LUCID}$ from the ratio between the number of non-colliding BCs and the total number of BCs. The two 
following methods in principle determine $\mu _{LUCID}$ from the ratio of the mean number of particles per BC and the mean number of particles 
per inelastic interaction, both seen by LUCID. Hit counting normally refers to particle counting, but where the counting capability of the 
detector is limited by its granularity.

The main requirements of the corresponding detector system are, an acceptance to minimum bias events, sufficient time resolution to measure 
individual BCs and being capable of counting particles. For this purpose ATLAS has chosen the LUCID detector which consists of aluminum tubes 
filled with $C_4F_{10}$ surrounding the beam-pipe and which are pointing at the ATLAS IP. The Cerenkov light emitted by a transversing particle 
is reflected down the tube and read-out by PMTs. The signal amplitude from the PMTs can be used to distinguish the number of particles per tube 
and the fast time response allows to measure individual BCs. A small scale LUCID, dedicated purely to luminosity monitoring, has been validated 
in testbeams and will be installed for the start up of the LHC. Based on the performance of the initial detector an optimized upgrade, including 
a large number of tubes, is foreseen to be installed at the same time as the upgrade of the LHC for the nominal luminosity of 
${\cal L} = 10^{34} ~cm^{-2}s^{-1}$.

For the luminosity measurement, the general calibration strategy of LUCID is to run in parallel with an absolute measurement of the luminosity 
at the ATLAS IP. Initially this will most likely be obtained from the LHC machine parameters with an expected precision of about 10-15\%. This 
will hopefully be improved in the medium term by studies of well known physics processes, like for example $W$ or $Z$ production as discussed in 
\cite{wz_counting} at this conference. When the ALFA measurement is available this will be the main reference for calibration. In this scenario 
the parallel measurement of $\mu _{LUCID}$ and $L$ will be made at optimal conditions for the absolute method (which provides $L$). The calibration 
constant, containing $\sigma _{inel}$ and $\varepsilon _{LUCID}$, can then be determined, allowing the LUCID measurement to directly provide the 
luminosity at different conditions.

\section{The ZDC system}

The third forward system in ATLAS is the {\it zero degree calorimeter}, which will measure neutral particles at a $0^{\circ}$ polar angle. The 
ZDC has a central role in the ATLAS heavy ion (HI) program where it is used to measure the centrality of the collisions, the luminosity as well 
as to provide triggers. It will, however, also be of importance both in the $pp$ program as described below and for accelerator tuning where it 
can be used to determine the location of the IP and the beam crossing angle.

The ATLAS ZDC consists of six tungsten/quartz calorimeter modules where the light from the quartz fibers is read-out by PMTs. In addition the ZDC 
is equipped with horizontal quartz rods, parallel to the beam, in order to determine the location of the showers in the plane perpendicular to the 
beam. The ZDC has been extensively tested and will be installed at the start up of the LHC. An upgrade is foreseen after about one year of running 
when additional space in the TAN absorber will become available.

\begin{wrapfigure}{r}{0.62\columnwidth}
\centerline{\includegraphics[width=0.60\columnwidth]{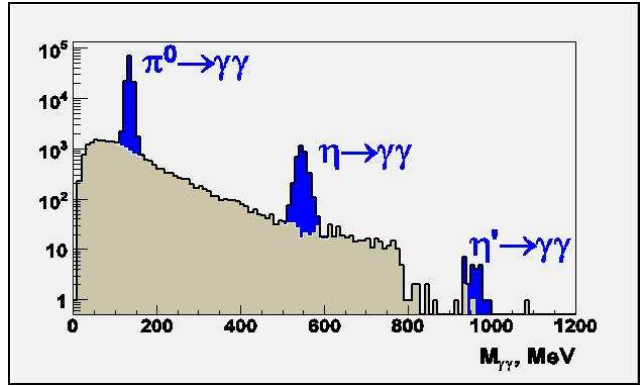}}
\caption{Simulated invariant mass spectrum as measured by the ZDC.}\label{Fig:ZDC}
\end{wrapfigure}
In the HI runs the main purpose of the ZDC is to measure the spectator neutrons. These are remnants of the collision and provides information about 
both the magnitude and direction of the impact parameter. In addition, the ZDC have close to a 100\% acceptance for HI collisions and together with 
the well known cross section of neutral particles at a zero degree angle the luminosity can be determined to a precision better than 5\%. It was 
also shown at RHIC that neutron tagging with the ZDC was essential to design a low rate trigger for ultra-peripheral events.

In the ATLAS $pp$ program the ZDC will mainly be used to study forward particle production. Figure \ref{Fig:ZDC} shows a simulated invariant mass 
spectrum as measured by the ZDC. Several meson peaks are clearly visible and also other mesons and baryons can be reconstructed. The cross section 
measurements of particles in the forward direction at the LHC energy is of interest for several applications. For example the measurement is of large 
interest to the high energy cosmic ray community where the information is required to properly model air showers from high energy protons entering the 
atmosphere, where the proton energy at the LHC, $E_{lab} = 10^{17} ~eV$, is just below the {\it knee} in the cosmic ray energy spectrum. In addition 
the ZDC will add to the overall hermeticity of ATLAS which will be useful to suppress background in diffractive studies. 


\begin{footnotesize}


\end{footnotesize}


\end{document}